\documentclass[%
aip,
amsmath,amssymb,
reprint,%
]{revtex4-1}

\usepackage{graphicx}
\usepackage{dcolumn}
\usepackage{bm}

\usepackage[utf8]{inputenc}
\usepackage[T1]{fontenc}
\usepackage{mathptmx}
\usepackage{etoolbox}
\usepackage{color}
\usepackage[dvipsnames]{xcolor}
\usepackage[normalem]{ulem}

\makeatletter
\def\@email#1#2{%
	\endgroup
	\patchcmd{\titleblock@produce}
	{\frontmatter@RRAPformat}
	{\frontmatter@RRAPformat{\produce@RRAP{*#1\href{mailto:#2}{#2}}}\frontmatter@RRAPformat}
	{}{}
}%
\makeatother

\usepackage{siunitx}

\newcommand{\Pe}{\mathrm{Pe}}
\newcommand{\LB}[1]{{\color{black} #1}}
\newcommand{\LBnew}[1]{{\color{black} #1}}
\newcommand{\PR}[1]{{\color{black} #1}}

\graphicspath{{figures/}}

\usepackage{placeins}
\usepackage{amsmath}

\newcommand{\rev}[1]{{\color{black} {#1}}}

\begin{document}
	
	\preprint{AIP/123-QED}

	\title{Nanofluidics at the crossroads}

	\author{Paul Robin}%
	\author{Lyd\'eric Bocquet}
	\email{lyderic.bocquet@ens.fr}
	\affiliation{Laboratoire de Physique de l'\'Ecole Normale Sup\'erieure, ENS, Universit\'e PSL, CNRS, Sorbonne Universit\'e, Universit\'e Paris-Cit\'e, Paris, France}
	\date{\today}
	
	\begin{abstract}
		Nanofluidics, the field interested in flows at the smallest scales, has grown at a fast pace, reaching an ever finer control of fluidic and ionic transport at the molecular level. Still, artificial pores are far from reaching the wealth of functionalities of biological channels that regulate sensory detection, biological transport and neurostransmission -- all while operating at energies comparable to thermal noise. Here, we argue that artificial ionic machines \LB{can be designed by harnessing} the entire wealth of phenomena available \LB{at the nanoscales} and exploiting techniques developped in various fields of physics. As they are generally based on solid-state nanopores, rather than soft membranes and proteins, they should in particular aim at taking advantage of their specific properties such as their electronic structure or their ability to interact with light. These observations call for the design of new ways of probing nanofluidic systems.
Nanofluidics is now at the crossroads, \LB{there are new avenues to build complex ionic machines and this may allow to develop new functionalities inspired by Nature.} 
		
		
	\end{abstract}
	
	\maketitle
	
	\section{Introduction}
	
	Nanofluidic transport is ubiquitous in Nature, and living organisms rely on membranes with complex properties to interact with their environment. They are notably involved in sensory detection (audition, touch, thermoception), osmotic regulation and neurotransmission \cite{hodgkin1952quantitative,coste2010piezo1,lumpkin2010review,story2003anktm1,kung2010mechanosensitive,perozo2002physical,viana2002specificity}. Similarly, cells use active ion pumps to create transmembrane concentration gradients \cite{gouaux2005principles} and decouple their chemical composition from that of the extracellular medium -- in biological terms, thermodynamic equilibrium is as good as death.
	
	In these few examples, the functions of biological membranes are governed by specific ion channels that react to certain external stimuli, opening or closing depending on conditions \cite{hille1992ionic,gerstner2002spiking}. They have long \rev{drawn considerable attention, as biomimetic membranes with similar properties would find use in multiple technological applications, from water desalination to the production of hydrogen or the development of iontronic machines.}
	
	The properties of biological systems, however, emerge from a subtle balance of `soft' processes that operate over energy scales up to a few times the thermal agitation $k_B T$, as determined by the structural and chemical properties of proteins of ion channels. The resulting excitability of biological membranes is in sharp constrast with `hard' condensed matter, where existing electronic systems are optimized to work well-away from thermal noise. Bridging the gap between these two worlds and creating bio-inspired machines therefore requires to find particular avenues where electronic processes could interact with soft matter.
	
	Over the last decade, research in nanofluidics has allowed to develop artificial fluidic systems with sizes reaching the molecular limit using a variety of materials and geometries \cite{kavokine2021fluids}: carbon or boron nitride nanotubes\cite{siria2013giant,secchi2016massive}, decorated pores in different types of membranes \cite{jain2015heterogeneous,acar2019biomimetic}, layered materials (graphene, MoS$_2$, graphene oxide) \cite{radha2016molecular,joshi2014precise}, etc. Yet, these devices are still greatly limited in terms of functionalities and the understanding of the physical phenomena occuring at the nanoscale, \LB{as well as their upscaling for potential applications}.
	
	Major challenges include the precise description of electrolytes confined in nanoscale devices: measurable quantities typically consist of ionic current under various external stimuli (voltage, pressure, concentration drop), but these macroscopic fluxes do not give precise information on the underlying transport processes. As a result, infering the properties of a nanofluidic channel from current-voltage (or current-pressure) characteristics requires to disentangle the effects of surface charges, hydrodynamic slippage, diffusioosmosis, surface adsorption, etc. Even when individual phenomena are well-understood on their own, existing models generally include multiple fitting parameters, making interpretation and comparison with experiments rather complex \cite{bocquet2010nanofluidics,kavokine2021fluids}.
	
	\rev{In addition, nanometric confinement has been reported to affect the properties of water itself, with strong modification of its thermodynamic\cite{algara2015square,yang2020capillary,kapil2022first} and electrodynamic\cite{fumagalli2018anomalously,artemov2021electrodynamics} properties. Likewise, nanoscale water transport, and in particular solid-liquid friction\cite{secchi2016massive,schlaich2017hydration,kavokine2022fluctuation}, is also still poorly understood. Finer understanding of the structure of confined water and ions would have a far-reaching impact on water desalination\cite{werber2016materials} and osmotic energy harvesting\cite{siria2017new}.} 
	
	Consequently, the field of nanofluidics would greatly benefit from novel techniques that would allow a more direct access to the microscopic properties of nanochannels through new sets of observables. In particular, finding ways of imaging nanoscale transport would constitute a quantum leap for nanofluidics.
	
	Notably, such techniques would be useful to make better use of possible couplings between nanoscale flows and solid walls. Recent advances have opened the avenue to the design of channels with carefully engineered electronic properties \cite{kavokine2022fluctuation,lizee2023strong,coquinot2023quantum}. Fine understanding of interactions between liquids and surrounding solid surfaces -- beyond the description of the interface as a boundary condition for water flows and electrostatic potential -- can only be achieved with specific techniques developped for problems in condensed matter. Such advances are possible by, for example, coupling nanofluidic measurement cells to a microscopy or spectroscopy apparatus.
	
	Another hurdle is the difficulty to correctly interface nanofluidic devices. In living cells, it is common for multiple pores and channels to be involved in a single process, such as the emission of action potentials in neurons and their stable propagation along (sometimes meter-long) nerve fibers \cite{hodgkin1952quantitative,gerstner2002spiking}. The overall process is governed by the repeated opening and closing of voltage-gated ion channels, as famously described by Hodgkin and Huxley. These biological examples have triggered hopes of creating artificial iontronic machines, that could perform specific tasks or carry out computations using ions as charge carriers -- as opposed to electrons in modern technology \cite{chun2015iontronics,robin_modeling_2021,robin2023long}. However, the leap from existing devices, that usually consist in single trans-membrane channels, to full networks of connected nanofluidic systems has yet to be overcome. The main limiting factors include the difficulty to probe systems locally with multiple electrodes, and geometrical constraints on fluidic cells.
	
	This Perspective is organized as follows. In Section II, we analyze the potentialities and limitations of bio-inspired ionic machines, notably through the example of pressure-sensitive channels. By comparing the energy levels over which biological and artificial systems operate, we show that achieving the same proficiency as biological systems requires to exploit novel ideas, such as \LB{going far from equilibrium or} using the electronic properties of channel's walls to impact water flows. We then review recent advances in that direction, and suggest potential applications. In Section III, we explore how nanofluidic systems could be probed and characterized, \LB{beyond} traditional conductance measurements. We notably focus on novel imagery techniques, and on how different types of external inputs (notably optical) could be used as probes. In Section IV, we address possible avenues in the design of more complex nanofluidic \LB{architectures}. In particular, we discuss the need for integrated systems, that could be connected and coupled to one another, \LB{while} probed locally. 
	
	\section{Ionic machines: nanofluidics at the edge of thermal noise}
	
	\subsection{Nanochannels as osmotic regulators and pressure sensors}
	
	One of the earliest challenges faced by unicellular organisms has been to equilibrate their osmotic pressure with that of their environment. Life emerged in oceans, whose water contains various electrolyes with total concentration around 1\,M, while a concentration difference of as little as 20\,mM (osmotic pressure of 0.5\,bar) is enough to rupture bilipid layers of cells \cite{kung2010mechanosensitive}. In reaction, organisms developped mechanosensitive ion channels -- pores that open or close depending on the pressure drop between the interior of the cell and the surrounding medium, releasing osmolytes whenever their cellular concentration becomes too great.
	
	Well-known examples include the mechanosensitive channels of \textit{E. coli}, known as MscS (small) and MscL (large) \cite{martinac1987pressure,kung2010mechanosensitive}. The latter consist in pores of radius around 3\,nm and length 10\,nm that expand by around 50\% for an osmotic pressure drop of more than 0.1\,bar. It is thought that the pressure drop induces a tension increase in the membrane of the cell, causing a confomation change in the MscL protein (Fig. 1A). This structural modification increases the cross-section -- and so the conductance -- of the pore.
	
	In a recent experiment by Davis and coworkers, an artificial pore with similar design was studied \cite{davis2020pressure}. The elastic membrane supporting the pore could stretch under a pressure drop, enlarging the pore and increasing its conductance (Fig. 1B). However, despite reporting the same mechanism and phenomenology, this type of channel only displayed a conductance increase of around 0.5\% under a pressure drop of 1\,bar, ie. two orders of magnitude below what is found in Nature.
	
	Such a discrepancy can be readily understood. Let us consider a pore drilled into a circular elastic membrane of Young's modulus $E$ and thickness $h$. Assuming the membrane is subject to a pressure drop $\Delta P$, the relative enlargement of the pore will be\cite{vlassak1992new}
	\begin{equation}
		\epsilon \sim \left(\frac{R \Delta P}{h E}\right)^{2/3}
	\end{equation}
	where $R$ is the membrane's radius. Biological membranes typically have $h \sim 1\,$nm and $R\sim1\, \si{\micro m}$ -- approximatively the limits of what can be done artificially. Preventing osmotic shocks imposes to obtain $\epsilon \sim 1$ for $\Delta P \sim 0.1\,$bar, yielding $E \sim 10 \,$MPa, comparable to biological membranes but orders of magnitude below standard artificial \LB{(solid-state)} materials.
	
	From a broader perspective, one can estimate the minimum detectable pressure regardless of the detailed structure of the pore through an energy estimate. Let us assume that the pore can be in either of two states -- closed or open -- separated by an energy barrier $\Delta E$, with the closed state being the more stable one. Direct dimensional analysis suggests that the state change will occur for $\Delta P \sim \Delta E/ \mathcal V$, where $\mathcal V$ is the pore's volume -- irrespective of exact mechanism. Typical biological pores like MscL have $\mathcal V \sim 10^{-24}\,$m$^3$ and conformation changes in proteins occur when a couple of hydrogen (or van der Waals) bonds are broken, ie. $\Delta E \sim 0.1\,$eV. This is again consistent with a detection threshold of $\Delta P \sim 0.1\,$bar.
	
	\begin{figure*}
		\centering
		\includegraphics[width=\linewidth]{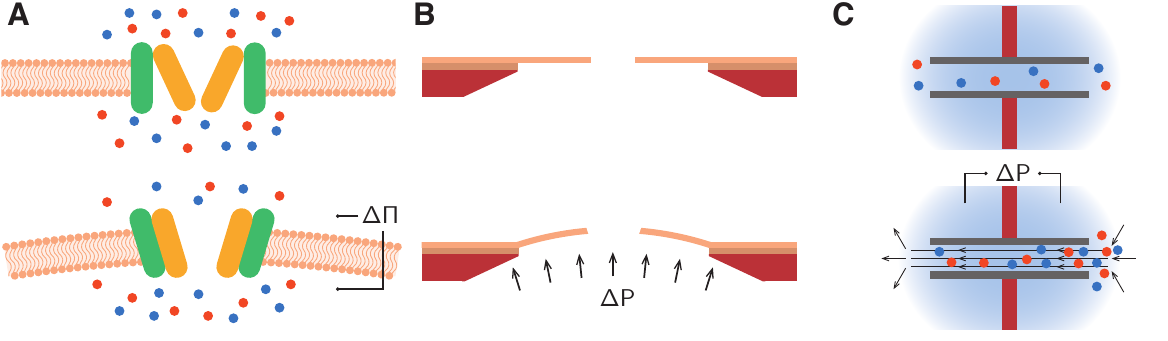}
		\caption{\textbf{Mechanosensitive channels.} \textbf{A} Example of biological ion channel: osmoregulation mechanism of MscL of E. coli. Top pannel: at osmotic equilibrium, the pore is in a closed state with low conductance.  Bottom pannel: when turgor builds up during an osmotic downshock, the tension of the lipidic membrane increases and causes a structural change of the pore's protein, allowing ions to flow through it. \textbf{B} Bio-inspired pressure sensitivity in an artificial membrane (adapated from \cite{davis2020pressure}). A nanometric pore in an elastic membrane may expand under an applied pressure. This geometrical deformation leads to an increased conductance. \textbf{C} Pressure sensitivity in carbon nanotubes. An external pressure drop applied across the nanotube creates a converging water flow, leading to an accumulation of ions inside the channel and in turn to a conductance increase.}
	\end{figure*}
	
	A similar discussion can be made for voltage-gated ion channels, found for example in neurons. These channels open or close depending on the voltage drop across the cellular membrane. This switching behaviour is thought to be driven by charged groups on the pore's protein: under an applied electric field, they reorient and thus impact the structure of the protein \cite{catterall1995structure}. A voltage drop of $\Delta V \sim 0.01-0.1\,$V is needed to overcome the energy barrier $\Delta E\sim 0.1-0.01\,$eV, again consistent with measurements on biological systems. Importantly, this voltage corresponds through Nersnt's equation to variations of concentration of order $1-10$ across the membranes of cells, allowing biological systems to easily detect chemical fluxes. For example, incremental changes in calcium concentration are known to play a role in synaptic plasticity \cite{gerstner2002spiking}. Energy barriers much larger than $k_BT$ would prevent the detection of subtle concentration changes: operating at the edge of thermal noise gives biological systems unique excitability properties, paradoxically resulting in higher accuracy for signal detection.
	
	The above analysis leads to a double-edged conclusion. It shows that new functionalities can emerge when probing nanofluidic devices with different stimuli (eg. voltage, pressure, concentration drops), \LB{provided} these are strong enough to affect the surrounding walls. There is, however, a fundamental limitation in the extent to which we can reproduce these effects in artificial systems: biological channels evolved over billions of years, and are tailored to interact with soft matter systems where the relevant energy scale is of the order of $1-10 \, \si{kJ/mol}$ (or $0.01-0.1\,$eV, or $0.5-5 k_BT$). We can reproduce this interplay between ion transport and changes in the pore's state; however this would require to use materials that can undergo changes in the same range of energies. Mechanical changes are unlikely to suffice, as typical materials have a Young modulus in the order of $1\,$GPa or above, though one could use pores ``decorated" by soft polymers, for example.
	
	Consequently, the design of e.g. artificial pressure-sensitive channels requires to exploit mechanisms other than the one used by living organisms, as we cannot directly reproduce the versatility of proteins. In other words, we should aim at replicating the function, but not the mechanism: we are unlikely to outperform biological systems evolved over billions of years at their own game. 
	
	Illustrating this principle, Marcotte et al. report that narrow carbon nanotubes can display mechano-sensitivity, due to the low friction of water on carbon surfaces (Fig. 1C) \cite{marcotte2020mechanically}. An external pressure drop applied on the tube causes ions to accumulate inside the system as they are dragged by the pressure-driven water flow. This results in an increase in salt concentration inside the channel, which can be estimated by equilibrating the number of ions advected by water with the number that leave the tube through diffusion. Imposing that the overall conductance increase $\Delta G$ must be an even function of applied pressure $\Delta P$, it can be estimated to scale like (for low applied pressure):
	\begin{equation}
		\Delta G \sim G_0 \Pe^2 \quad ; \quad \Pe = \frac{K}{D}\Delta P
	\end{equation}
	with $G_0$ the conductance in absence of a pressure drop, $\Pe$ the Péclet number, $K$ the permeance of the nanotube \LB{and} $D$ the diffusion coefficient of ions. In most instances in nanofluidics, the Péclet number is small due to the nanometric sizes of the considered channel (with typically $\Pe \sim 10^{-2}$), and consequently the mechano-sensitive effect is negligible. This picture changes, however, for \LB{channels exhibiting} strong hydrodynamic slippage on \LB{their} surface. Slippage increases considerably the permeance of small channels, multiplying the Péclet number by a factor of the order of $b/R$, with $b$ the slip length. In small carbon nanotubes, where the slip length can reach $b \sim 100\,$nm \cite{secchi2016massive}, one can typically reach $\Pe \sim 1-10$, resulting in a conductance increase of around 30\% under 0.5\,bar, close to the performance of biological systems. The mechanism at works is, however, completely different, relying on hydrodynamic slippage rather than mechanical deformation of soft materials.

	\subsection{Electronic effects in solid-liquid friction}
	
	\begin{figure*}
		\centering
		\includegraphics[width=\linewidth]{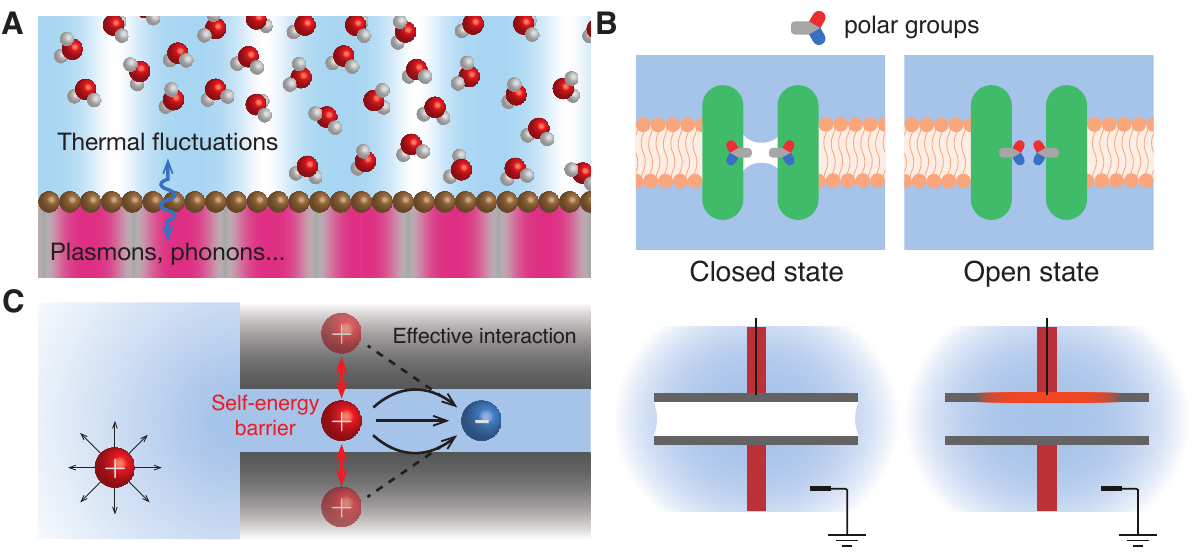}
		\caption{\textbf{Electro-hydrodynamic couplings.} \textbf{A} Influence of walls' electronic structure on solid-liquid friction. On atomically smooth surfaces like graphite, interfacial friction is expected to be dominated by the dissipation of momentum caused by electronic modes that resonate with water fluctuation modes. For example, the plasmon mode of graphite has an energy close to $k_B T$ and can thus be excited by thermal fluctuations of water. \textbf{B} Channels with tuneable wettability. Top pannel: some biological ion channels are known to switch between low- and high-conductance states by changing their wettability. External parameters like pressure or potential can change the structure of the pore's protein, causing polar groups to be either hidden or accessible. When polar groups are hidden, water is repelled from the pore, block the passage of ions. Bottom pannel: Idea of an artificial channel with tuneable wettability. At rest, the channel repels water and is non-conducting. A local electrode allows to modify the potential of channel's walls. At high enough potential, water permeates the channel through electrowetting. \textbf{C} Interaction confinement in nanometric channels. The presence of walls modify the electrostatic potential around ions, impacting the self-energy for an ion to enter the pore as well as ion-ion interactions. This process can be interpreted in terms of image charges, which magnitude and sign depend on the metallicity and the electonic structure of walls.}
	\end{figure*}
	
	The previous example illustrates how a careful choice of materials for building nanofluidic channels can allow to reproduce complex functionnalities. Hydrodynamic slippage is strongly dependent on the surface properties of wall materials. While in most situations it is essentially determined by the roughness of surfaces, \LB{coulombic} coupling between electronic modes of the solid and nearby water molecules can also play a role. When the frequency of electronic modes matches that of thermal fluctuations in water, momentum from the water flow can be dissipated in the form of conduction in the channel's walls. In the case of an atomically smooth surface (e.g. carbon nanotubes or \LB{graphite} surfaces), this \LB{quantum} contribution can even dominate all other sources of solid-liquid friction (Fig. 2A) \cite{kavokine2022fluctuation}.
	
	Interestingly, quantum friction is relevant for materials with electronic modes \LB{with energies on} the order of $0.1\,$eV or below, comparable to the Debye peak of water. This analysis shows that, while `hard' artificial materials usually cannot couple to `soft' excitations through mecanical deformations and structural changes (due to their high Young modulus), there is still hope to recreate the functionalities of protein-based membranes through electronic coupling. 
	
	In particular, carbon-based materials are promising candidates to achieve this coupling. Graphite, for example, displays a \LB{dispersionless,} low-energy plasmon mode at energies close to $k_B T$ \cite{portail1999dynamical}. Exploiting its properties could open the way for the control of nanoscale water flows using the electronic properties of surfaces, e.g. through doping \LB{or gating}. More generally, solid-liquid friction can be seen as a coupling parameter between the physics of liquids -- flows of water or ions driven by voltage, pressure or concentrations drops, for example -- and that of condensed matter. Solids display a wealth of excitation modes that can be tuned through careful engineering: for example, interfacial liquid flows have been shown to excite phonon modes within the solids \LB{whose properties can be tuned by proper engineering of the confining surfaces} \cite{coquinot2023quantum,lizee2023strong}.
	
	\subsection{Gated nanofluidic transport}
	
		\begin{figure*}
		\centering
		\includegraphics[width=\linewidth]{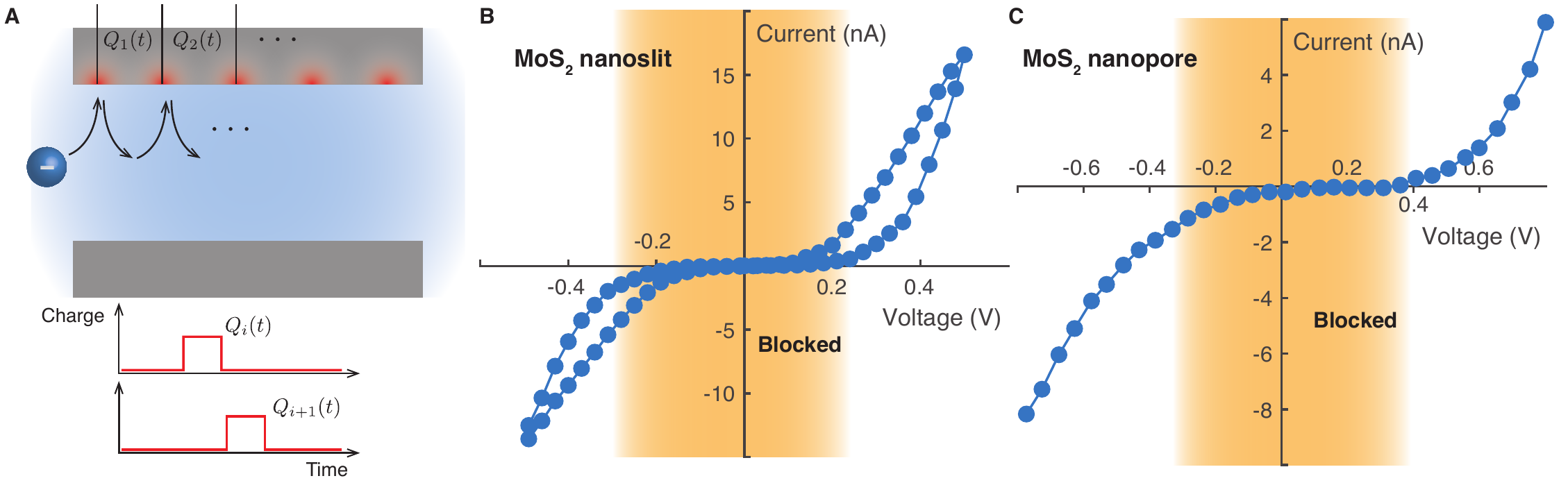}
		\caption{\textbf{Voltage-gated nanofluidic channels.} \textbf{A} Schematic design of a single-ion pump, as suggested by ref. \cite{kavokine2019ionic}. Several electrodes are placed along the surface of a nanotube, each bearing a local charge, with which an ion of opposite sign can pair up. Varying these charges in time allows to drag a single ion across the channel, creating a pumping mechanism. \textbf{B} Voltage gating in a 2D MoS$_2$ nanoslit filled with KCl $3 \, \si{M}$ (adapted from ref.~\cite{robin2023long}). Conductance strongly depends on applied voltage, and displays a hysteresis effect over long time scale (here, a voltage sweep of amplitude $0.5\,$V and period 40 min was imposed on the channel). This corresponds to the definition of a memristor. \textbf{C} Voltage gating in a MoS$_2$ nanopore filled with KCl 2\,M (adapted from ref.~\cite{feng2016observation}).}
	\end{figure*}
	
	One key aspect of biological channels is their ability to block or allow passage of water or ions upon certain precise conditions. This process is key to maintaining the composition of the intracellular medium, which would otherwise be dictated by the external environment. As a result, gated nanochannels would serve as elementary building blocks for biomimetic applications.
	
	Gating processes rely on tunable energy barriers that prevent water or ions from entering biological pores. These barriers must at the same time be high enough to result in good rejection rates (such that $e^{\Delta E/ k_B T} \ll 1$) and easily counterbalanced by other biological processes like protein conformation changes; overall, this sets again $\Delta E \sim 5 \, k_B T$, allowing for rejection rates greater than $99\%$, while being comparable to the breaking of a few van der Waals bounds. This energy scale means that random switching between physical states (say from the state with no ions inside the channel to a state with one ion inside) are unlikely enough to guarantee the system's robustness, while preserving excitability -- i.e., the system's capability to react to even weak stimuli. When greater levels of robustness are required, a fruitful strategy is generally to rely on $N$ interacting channels with the same energy barrier $\Delta E \sim 5 \, k_B T$, rather than to rely on a single one with energy barrier $N \Delta E$. In both cases, the rate of random switching between states will scale like $e^{-N\Delta E/k_B T}$; however, higher energy barriers also mean much lower excitability, as fewer external stimuli will be strong enough to trigger a response. The latter case would yet correspond to most artificial systems that rely on electronic or mechanical processes typically have energy barriers of the order of $\Delta E \sim 1 \, \si{eV} = 40 \, k_B T$, so that $e^{-\Delta E/k_B T}\sim 10^{-18}$: the gain in robustness is outweighted by the loss in excitability. In contrast, chemotactic bacteria like E. coli are able to detect concentration changes of a few nanomolars (equivalent to a few molecules per cell volume)\cite{mao2003sensitive}; this very high degree of precision is attributed to collaborative behaviours in clusters of several similar chemoreceptors\cite{briegel2009universal,sourjik2012responding}.
	
	Modifications in wetting of biological channels, for example, are known to play an important role in their ability to switch between conducting and non-conducting states (by attracting or repelling water from the pores) \cite{yazdani2020hydrophobic}. Studies suggest that these modifications, in some cases, are controlled by conformation changes of the pores' proteins resulting in hiding or exposing hydrophilic groups (Fig. 2B) \cite{anishkin2010hydration}. 
	
	Reproducing the same phenomenology in artificial pores \LB{is challenging} due to the reasons presented above. However, it could {e.g.} be imitated by modifying the surface properties of walls. The contact angle of hydrophobic surfaces is known to decrease with their electrical potential due to the development of an electrostatic double layer \cite{mugele2005electrowetting}. Assuming the double layer has a molecular size, the resulting capacitance is of the order of $C \sim 10\, \si{\micro \farad / \centi \meter \squared}$.  The modification in surface tension under $V \sim 1\,$V can be estimated to reach $C V^2 \sim 10^{-1} \, \si{\joule / \meter \squared}$. This would imply that electrowetting could counteract the surface tension of water. If the potential of a small hydrophobic pore could be tuned with a local \LB{gating} electrode, one could then imagine to gate the passage of water through the pore (Fig. 2B), \LB{via an external gate potential} \cite{jiang2011charge}. \LB{While the gating configuration has not been achieved to our knowledge in this context, a voltage bias across a nanopore was shown to lead to similar results \cite{powell2011electric}, in line with previous numerical and theoretical predictions \cite{dzubiella2005electric}}.
	
	Similarly, the nature of walls have been shown to impact the behaviour of ions confined in nanochannels. As known since the pioneer work of Parsegian\cite{parsegian1969energy}, ions experience strong energy barriers when crossing lipidic membranes, due to the dielectric contrast between water and lipid walls. Electric field lines created by ions are confined within the channel: consequently, a confined ion will possess a higher electrostatic self-energy compared to an ion in bulk water (Fig. 2C). This excess electrostatic energy can be thought as the interaction between ions and the image charges in the insulating lipid walls: a dissolved charge $q$ next to planar insulator will create an image charge with magnitude\cite{kavokine2022interaction}
	\begin{equation}
		q_\text{im} = \frac{\epsilon_w - \epsilon_m}{\epsilon_w+\epsilon_m}q
		\label{eqn:qim}
	\end{equation}
	where $\epsilon_w$ is the dielectric constant of water, and $\epsilon_m$ that of the external medium. Because lipidic membranes have $\epsilon_m \sim 5$, the image charge will have the same sign as the original ion, resulting in a destabilizing interaction. Since the image charge is created symmetrically to the original particle, this energy barrier scales like\cite{robin2023ion}:
	\begin{equation}
		\Delta E \sim \alpha \frac{q^2}{\epsilon_0\epsilon_w R}
		\label{eqn:deltaE}
	\end{equation}
	where $R$ is the size of the channel and $\alpha$ a material- and geometry-dependent screening factor of order unity. Consequently, wall materials can be expected to play a dominant role in the properties of solvated ions in channels of size comparable to the Bjerrum length $\ell_B = e^2/4 \pi \epsilon_0 \epsilon_w k_B T$. At room temperature, one obtains $\ell_B \sim 1\,$nm. However, \rev{dielectric properties of water are known to strongly deviate from that of bulk water under nanometric confinement\cite{fumagalli2018anomalously}. Water notably becomes anisotropic, with a dielectric constant $\epsilon_\perp \sim 2$ in the direction of confinement -- while other directions remain unaffected, with $\epsilon_\parallel \sim 80$. Overall, one should typically replace $\epsilon_w$ by $\sqrt{\epsilon_\perp \epsilon_\parallel}$ in Eqs.~\eqref{eqn:qim} and \eqref{eqn:deltaE} \cite{kavokine2019ionic,robin_modeling_2021}. Hence, modifications to the dielectric properties of water at the nanoscale generally result in stronger electrostatic effects.}
	
	Similarly, the electrostatic interaction between solvated ions of opposite charge will be strengthened in very small channels, again because of interactions with image charges \cite{kavokine2022interaction}. This phenomenon, termed interaction confinement, can ultimately lead to the formation of tightly bound ion pairs, or Bjerrum pairs. As ion pairs are neutral, pairing strongly impacts the conductivity of nanochannels and has been shown to result in non-linear and hysteretic ion transport. 
	
	Quantitatively, one can estimate that the energy needed to break an ion pair is comparable to the self energy of the two resulting free ions, or $\Delta E \sim k_B T \times \alpha \ell_B/R$. To fix ideas, monovalent ions entering a channel with $R \sim 1\,\si{nm}$ and $\epsilon_m = 2$ will experience an energy barrier $\Delta E = 5 \, k_B T$. In other words, interaction confinement will reduce conductance by $99\%$ due to the formation of ion pairs. This barrier can be lifted through the application of an external voltage drop strong enough to break the pairs, so that conduction is restored -- a process known as the second Wien effect \cite{onsager1934deviations,kaiser2013onsager}. Theoretical modelling of the Wien effect has allowed finer understanding of the underlying mechanisms. In particular, this phenomenon replicates the physics of voltage-gated ion channels, and its modelization has opened the way for the design of solid-state channels with key properties. Examples include a single-ion pump (see Fig.~3\textbf{A}) \cite{kavokine2019ionic} and ionic memristors (see Fig.~3\textbf{B}) \cite{robin_modeling_2021}. On the experimental side, voltage-gating was observed in both short pores and 2D slit-like channels (see Fig.~3\textbf{B} and \textbf{C}) \cite{feng2016observation,robin2023long}. In the latter case, the dynamics of ionic pairing and pair breaking due to Wien effect was found to result in hour-long hysteresis in ionic conduction. The conductance of both nanopores and nanoslit was found to vary by a factor around 50 when applying a voltage drop $\Delta V \sim 0.5 \,$V, \LB{consistent} with the previous interpretations of voltage gating in terms of interaction confinement and ion pairs. \PR{This combination of non-linear response and memory is at the source of the ionic memristor \cite{robin2023long,xiong2023neuromorphic}. In electronics, memristors are used as solid-state equivalents of biological synapses due to their tunable conductance and their ability to store information over long timescales \cite{sebastian2020memory}. Here, this analogy is strengthened by the fact that nanofluidic memristors rely on ions as charge carriers, much like their biological counterparts. These elementary building blocks were also used to demonstrate a basic form of Hebbian learning\cite{robin2023long} -- a learning process based on causality detection.}
	
	Overall, it is possible to gate ion transport through nanochannels, either by applying a voltage of order $V \sim 0.1-1\,$V, or by externally modifying the screening factor $\alpha$ (e.g. through doping). This voltage-gating phenomenon is comparable to the one in biological channels -- both in function and in energy scales; however, the underlying physical principle is entirely different.
	
	\begin{figure*}
		\centering
		\includegraphics[width=\linewidth]{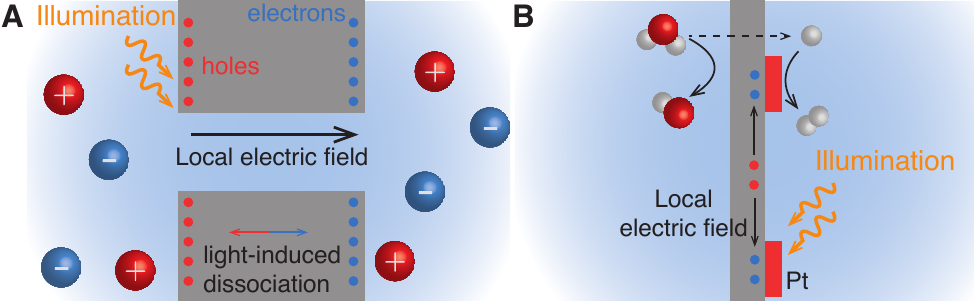}
		\caption{\textbf{Controlling ions with light.} \textbf{A} Optical ionic pump (adapted from \cite{xiao2019artificial}). A membrane of carbon nitride nanotubes is exposed to to light, which creates free electrons and holes inside the material. Electrons are repelled to the unilluminated side, creating a local electric field across the negatively-charged nanotubes. This process allows to selectively pump cations from one reservoir to another, even against strong concentration gradients. \textbf{B} Solar-enhanced water dissociation (adapted from \cite{cai2022photoaccelerated}). A graphene membrane coated with platinum particles is exposed to sunlight. As graphene is permeable to protons, but not to hydroxyde ions, a voltage drop applied across the system will create a net proton flux through the membrane. Sunlight generates electrons and positive holes, creating a local electric field driving protons towards platinum particles. This process allows to enhance dihydrogen production with sunlight.}
	\end{figure*}
	
	An advantage of artificial materials compared to lipidic membranes is that their electronic properties extend beyond that of simple insulators. In particular, all the effects discussed above (solid-liquid friction, electrostatic energy barriers and interaction confinement) have been predicted to be controllable through the tuning of the electronic structure of walls. Therefore, techniques like the doping of semi-conductors could allow to create new ionic machines based on these principles. Of particular importance are the energy scales of these phenomena: because they all fall in the range of $0.01-0.1\,$eV, or $0.5-5\,k_B T$, they could have a strong impact on water and ion transport and, to some extent, play the role of conformational changes in the proteins of biological pores. For example, \rev{one could imagine to block or allow the passage of ions through a nanochannel by tuning the electrostatic self-energy between, say, $k_BT$ and $5 k_B T$.}
	
	\subsection{Controlling ions with light}
	
	Another interesting property of electronic materials is their ability to interact with light, through the creation of conducting electrons and holes after the absorption of select wavelengths. This provides another avenue for the design of complex ionic machines. For example, Xiao and coworkers have reported the possibility to create an artificial ionic pump based on carbone nitride nanotube (CNNT) membrane (Fig.~4A) \cite{xiao2019artificial}. According to the authors, exposition to sunlight separates electrons and holes inside the membrane, generating a localized electric field that is able to move ions uphill concentration gradients.
	
	A similar idea was exploited by Cai and coworkers to enhance water dissociation using graphene membranes (Fig.~4B) \cite{lozada2018giant,cai2022photoaccelerated}. By imposing a voltage drop across a one-atom thick membrane, the authors were able to produce very high electric fields, enough to dissociate water molecules. As protons are able to diffuse through graphene layers, this process can be used to produce H$_2$ through platinum nanoparticles deposited on graphene. Here, light was used, not to create a local voltage drop across the membrane, but as a way to funnel protons towards Pt particles through the creation of an in-plane electric field: light is thought to create long-lived electrons in graphene, that are then attracted by Pt particles, and protons along with them.
	
	Lastly, one could imagine to gate ionic transport with light, exploiting the electrostatic effects detailed above. Light can generate local electic fields through the separation of electrons and positive holes; this process could serve as a way to break ionic pairs and therefore tune the conductivity of nanofluidic systems, providing a unique platform to locally control their properties.
	
	This coupling between light and nanofluidics will be the focus of the next section, where we will show how it can also be exploited to probe fluidic systems in new ways.

	\section{Probing nanoscale transport at the crossroads between nanofluidics and condensed matter}
	
	\subsection{The limitations of conductance measurements}
	
	A recurrent issue in nanofluidics is the lack of precise microscopic information on nanoscale devices. In most cases, measurable quantities consist in global flow rates through the system (fluxes of mass, charge or solute), which are the combined results of many physical processes occuring at the microscopic level. For instance, conductance can be governed by geometry, salt concentration, the presence of surface charges, hydrodynamic slippage, etc. This complexity prevents from linking observations to a direct cause (e.g., a conductance increase to a change in the surface charge). \rev{Worse, the high number of unknowns complexifies comparison with simulations or theoretical modelling}. One very common example of such unknown parameters is the concentration in ions inside a (sub)nanometric pore. Due to phenomena like dielectric contrast, or the deformation of solvation shells, salt concentration inside the pore may largely differ from that of the reservoirs -- even in absence of any surface charge -- due to energy barriers for ions to enter the pore. Since conductivity scales like the product of ionic mobility times the number of charge carriers, it does not allow direct access to salt concentration in situ: deviations from the bulk value could be attributed to modifications in mobility (due to e.g. hydrodynamic slippage or adsorption processes) as well as entry effects governed by electrostatics. 
	
	To take a more quantitative example, let us focus on conductance alone. At low concentrations (such that the Debye length $\lambda_D$ is larger than the channel's size), the conductance of a slit-like channel of dimensions $L \times w \times h$ bearing a surface charge 
	$\Sigma$ is expected to read\cite{kavokine2021fluids}
	\begin{equation}
		G = \frac{2 w h}{L} \frac{e^2 D}{k_B T}\sqrt{c_s^2 + \left(\frac{\Sigma}{h e}\right)^2} + \frac{wh}{L}\frac{\Sigma^2}{3 \eta} \left(1 + \frac{6 b}{h}\right)
	\end{equation}
	where $D/k_B T$ is the mobility of ions, $c_s$ the salt concentration in the reservoirs, $\eta$ the viscosity of water, $b$ the slip length and $e$ the elementary charge. The first term in the above equation corresponds to electrophoretic mobility (ions dragged by the electric field), and the second to electroosmosis (ions dragged by the water flow). Among these parameters, $\Sigma$ and $b$ are generally not known precisely; in particular, the surface charge is known to often depend on pH and salt concentration\cite{uematsu2018crossover,secchi2016scaling}, and the slip length to depend on the surface charge itself. In addition, the entry effects as detailed in the last section can potentially modify this whole picture for channels of size $1\,$nm or smaller: taking them into account would amount to replace $c_s$ by $c_s e^{- \Delta E/k_BT}$, with $\Delta E$ an unknown energy barrier.
	
	As a result, the characterization of nanofluidic devices often rely on the cross-checking of multiple measurements, say by measuring conductance in function of salt concentration to estimate the surface charge, and then modulating the latter through pH to \LB{infer} the slip length. In addition, many nanofluidic measurements are subject to repeatability or hysteresis issues -- in particular when pH is varied.
	
	Moreover, fluxes generally scale like a power of the system size -- for example, mass flow rate in a cylindrical tube of radius $R$ scales like $R^4$ in the absence of slippage -- meaning that they quickly become immeasurably small for systems close to $1\,$nm in size. As a consequence, it is often necessary to carry out measurements arrays of multiple nanofluidic channels to obtain a manageable signal-to-noise ratio. Due to the inherent variability of nanofluidic systems, this prevents fine and quantitative understanding of the phenomena at works, due to certain channels being \LB{possibly} blocked (introducing another unknown parameter in the number of active channels), or to interactions in entrance effects. This problem is particularly noticeable when measuring fluxes of mass or solute, due to the lack of general-purpose sensors with very high precision, but also affects conductance measurements for currents of the order of picoamperes.
	
	These combined aspects mean that for channels smaller than $1\,$nm, variations in measures by less than one order of magnitude are generally impossible to analyze. This calls for the design of new probes, that could allow a more direct access to the properties of nanochannels. In what follows, we review a few examples of such techniques, and discuss their relevance in future developments.
	
	\subsection{Imaging nanoscale transport}
	
	\begin{figure*}
		\centering
		\includegraphics[width=\linewidth]{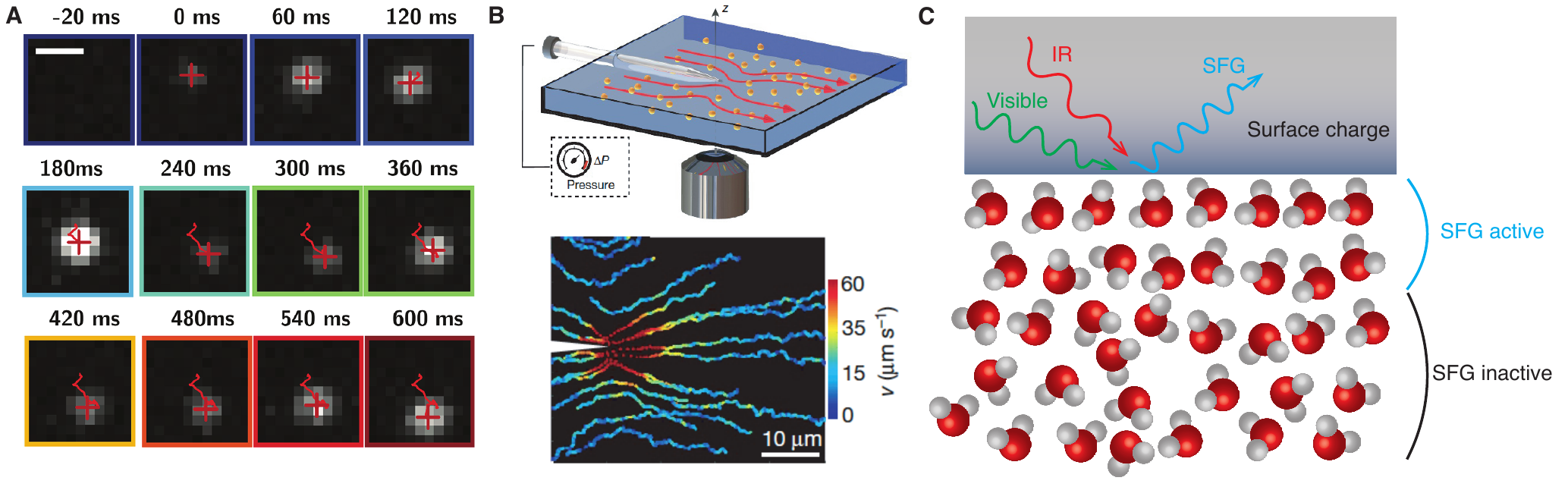}
		\caption{\textbf{Optical probes of nanofluidic transport.} \textbf{A} Diffusion of a single proton on light-emissive hBN defects (reproduced from \cite{comtet2020direct}). The position of a proton is tracked in time using superesolved microscopy. When protonated, natural borion vacencies of hBN become emissive. Scale bar represents $500\,\si{nm}$. \textbf{B} Measuring a flow rate through an individual carbon nanotube by tracking tracers trajectories (reproduced from \cite{secchi2016massive}). The jet emerging from the nanotube deforms the water flow locally and perturbs the trajectories of tracers. The magnitude of the perturbation is proportional to the velocity of the jet, allowing to measure the flow rate through the nanotube. \textbf{C} Principles of SFG spectroscopy to probe solid-liquid interfaces. The local orientational order of water molecules is perturbed near interfaces. This asymmetry in polarity can be detected by SFG spectroscopy, which measure electrical susceptibility by exposing the sample to two laser beams and measuring the output signal. This techniques notably allows to measure surfaces charges.}
	\end{figure*}
	
	\LB{One is basically blind at the nanoscales and }a long-term dream in nanofluidics has been to be able to image the physical processes at work in nanoscale systems. Considerable amount of progress has been made to describe the properties of liquid interfaces, and we believe that some of the techniques could be extended to probe confined liquids as well.
	
	A notable example is the imaging of single-proton diffusion on defects boron nitride surfaces: according to Comtet and coworkers, negatively charged boron vacancies are able to capture protons (Fig.~5A) \cite{comtet2020direct}. The protonated state can then be probed by laser exposition, promoting it to a radiative and detectable excited state. The authors were able to localize these excitated states with nanometric precision using super resolution microscopy, allowing to track the random motion of a single proton as it hops between boron vacancies. This imaging technique notably showed how proton diffusion is considerably slowed down near a chemically active surface. \rev{A similar technique was later used to image molecular diffusion}, this time in a confined environment consisting of a nanofluidic slit (height of $1-2\,$nm) \cite{ronceray2022liquid}. Such setups could allow direct access to the molecular dynamics inside nanochannels, particularly in cases where surfaces processes play an important role. The drastic reduction in the proton diffusion coefficient down to $D \sim 10^{-14}\, \si{\meter \squared \per \second}$ can be attributed to the high energetic cost for ions to escape defects. This description connects with the interpretation of hour-long memory effects observed in 2D nanochannels \cite{robin2023long}. Material defects in 2D materials have been thought to possibly trap ions on the channel's surface, slowing down their diffusion considerably. This process allows the development of ionic accumulations over long timescales, through the retentions of ions on the system's surface. Direct observation of this mechanism would represent a major advance.
	
	More generally, optical imaging techniques can also be used to measure transport quantities in indirect ways, but with a higher precision than conventional setups. This is particularly noticeable for the detection of mass fluxes. The most direct way to measure the flow rate of water across a nanochannel is through the use of a microbalance \cite{radha2016molecular}; however, this technique is typically limited to flow rates of the order of $10^{-12} \, \si{L  \per s}$. For an array of 100 slit-like channels with dimensions $1\,\si{\micro \meter} \times 1 \, \si{\nano \meter} \times 100 \, \si{\nano \meter}$ (or 100 nanotubes of radius $10\,$nm), this corresponds to an average water velocity of $0.1 \, \si{m \per s}$. For reference, this velocity is one order of magnitude higher than the flow created by a pressure drop of $\Delta P \sim 1 \,$bar across a nanotube of radius $10\,$nm and length $1\, \si{\micro \meter}$ (assuming no hydrodynamic slippage).
	
	Again, optical techniques can prove useful for measuring flow rates through individual systems. As flows emerging from a small channel into a reservoir tend to deflect nearby flowlines, it is possible to infer flow rates inside nanochannels from tracking tracer particles (Fig.~5B). For nanotubes, this technique allows to measure flows corresponding to a velocity of the order of $10^{-2}\, \si{m \per s}$ inside the channel, and thus gives unambiguous access to the tube's slip length \cite{secchi2016massive}.
	
	The two above examples show how optics can provide new observables to characterize the properties of nanofluidic systems. We believe that similar ideas could be put to fruition in other setups. For example, one could imagine to use a nanofluidic device to transport a dissolved reactant to a chemically active site: if the resulting reaction emits light, or is photo-activated, one could then measure the transport rate of the reactant across the nanofluidic system.
	
	Another promising technique for probing nanoscale physics is sum frequency generation (SFG) spectroscopy \cite{hunt1987observation}. This technique allows to measure the non-linear electrical susceptibility of interfaces, notably effectively tracking the orientational order of polar molecules -- most importantly, water -- at an interface. Critically, SFG spectroscopy allows to measure surface charges, as they cause important modification of water orientation in the first few molecular layers \cite{gonella2021water}. While, to the best of our knowledge, SFG spectroscopy has yet to be used in a nanofluidic environment, it could give access to time-resolved surface charge measurements. Again, the dynamics of surface charges is known to strongly impact ion transport, being associated with long relaxation times and hysteresis through the coupling to the Stern layer\cite{lis2014liquid,ober2021liquid}.
	
	Furthermore, the surface charged of systems immersed in water is regulated by a variety of processes, and notably through salt concentration and pH. In a complex environment, where for example several nanofluidic devices would be linked in a network such that one system collects the electrolyte fluxes of the others, one could imagine this surface equilibrium to evolve over time. In such conditions, direct access to the system's surface charge would be crucial.
	
	\subsection{Time-dependent observables}
	
	Another important limitation of traditionnal conductance measurements is their lack of information on the systems' dynamics. Biological pores often operate in a highly fluctuating environment; in neurons for instance, ion channels are typically subject to a potential change of $\sim 100\,$mV in $\sim 10 \,$ms \cite{hille1992ionic,hodgkin1952quantitative}. The resulting complex and hysteretic response of voltage-gated ion channels has been at the root of the Hodgkin-Huxley model that explains the generation and propagation of action potentials in nerve fibers, enabling communication between neurons. More generally, biological sensors are often adaptable and plastic, allowing for greater precision and learning. One of their main characteristics is to switch between open and closed states, reactively to changes in external conditions (as discussed in Section II). 
	
	Of particular importance is the phenomenon of adaptation, where biological sensors quickly become quiescent under a constant excitation. This property has been linked to specific physical processes occuring at the level of ion channels \cite{gerstner2002spiking,housley2013atp}. Adaptation allows biological organisms to broaden the working spectrum of their sensors (by resisting to even very strong excitations), or to focus on detecting changes, e.g. in chemical concentration of certain nutrients, rather than an absolute value. This effect is at the root of motion strategies of microbes in a process known as chemotaxis \cite{wadhams2004making}. Such complex behaviours rely on dynamical effects, where only time-varying stimulations are detected.
	
	While time-dependent phenomena are generally not probed for in nanofluidics, there are strong reasons to believe that they \LB{offer a lot of opportunities}. All processes that rely on the accumulation of ions inside nanochannels could display rich dynamics if excited on timescales comparable to the formation time of these accumulations. Notable examples include carbon nanotubes with pressure-modulated conductance, and all devices displaying ionic rectification, like nanofluidic diodes. In particular, these systems have been shown to behave as memristors -- electronic devices with hysteretic conductance -- when exposed to time-varying voltage\cite{robin_modeling_2021,robin2023long,xiong2023neuromorphic,noy2023nanofluidic}. A notable trait of such systems is to display a pinched loop in their current-voltage characteristic when exposed to a voltage sweep, as shown for example on Fig.~3\textbf{B}. This property can be interpreted as the system keeping track of the history of voltage it has been exposed to. In the emerging field of bioinspired computer science, such systems have been used as the electronic counterparts of biological synapses. Nanofluidic memristors would thus allow the design of biomimetic neuron networks. \LB{This is a first, key ingredient toward neuromorphic ionic computing \cite{noy2023nanofluidic}}.
	
	More generally, controlling ion transport in real time is a necessary step towards the development of iontronics -- the field interested in carrying out computations based on ions and fluidic devices, rather than electrons and solid-state systems. Indeed one would need to actuate nanofluidic systems with streams of inputs, and have them react to one another. We therefore believe that the interest of studying the temporal response of nanofluidic systems would be two-fold. It would provide additional information on nanoscale phenomena, and can allow to conceive new, dynamical ionic machines with greater complexity.
	
	Another way of probing the internal dynamics of a system is to look at its fluctuations. Nanofluidic signals (such as the ionic current across a nanopore under a constant voltage drop) are known to be correlated over timescales that extend far beyond reasonnable microscopic timescales\cite{secchi2016scaling,hooge19701f,smeets2008noise}. Biological systems are also subject to similar fluctuations\cite{wohnsland19971}, and their study can shed light on the microscopic state of biological channels -- e.g. the typical lifetime of ``open'' and ``closed'' states of the channels\cite{bezrukov2000examining}. More generally, artificial and biological pores alike are subject to $1/f$ pink noise, a poorly understandood and highly debated type of fluctuations described the empirical Hooge's law\cite{hooge19701f}
	\begin{equation}
		S_I(f) = \frac{\alpha I^2}{N f}
	\end{equation}
	where $S_I(f)$ is the power spectrum of ionic current under a constant voltage, $I$ the average current, $N$ the number of charge carriers inside the pore and $\alpha$ a dimensionless constant. While its origin would warrant additional scrutiny, this type of fluctuations could be used as a way to measure the number $N$ of ions present inside a nanochannel, and thus the local value of ionic concentration. \rev{Overall, using current fluctuations as a probe of local ion dynamics would require to pinpoint which aspects of pink noise is universal to ion transport and which ones can be attributed to specific nanoscale phenomena\cite{robin2023disentangling}.}
	
	\section{Discussion: towards a bio-inspired machinery}

\begin{figure*}
	\centering
	\includegraphics[width=0.9\linewidth]{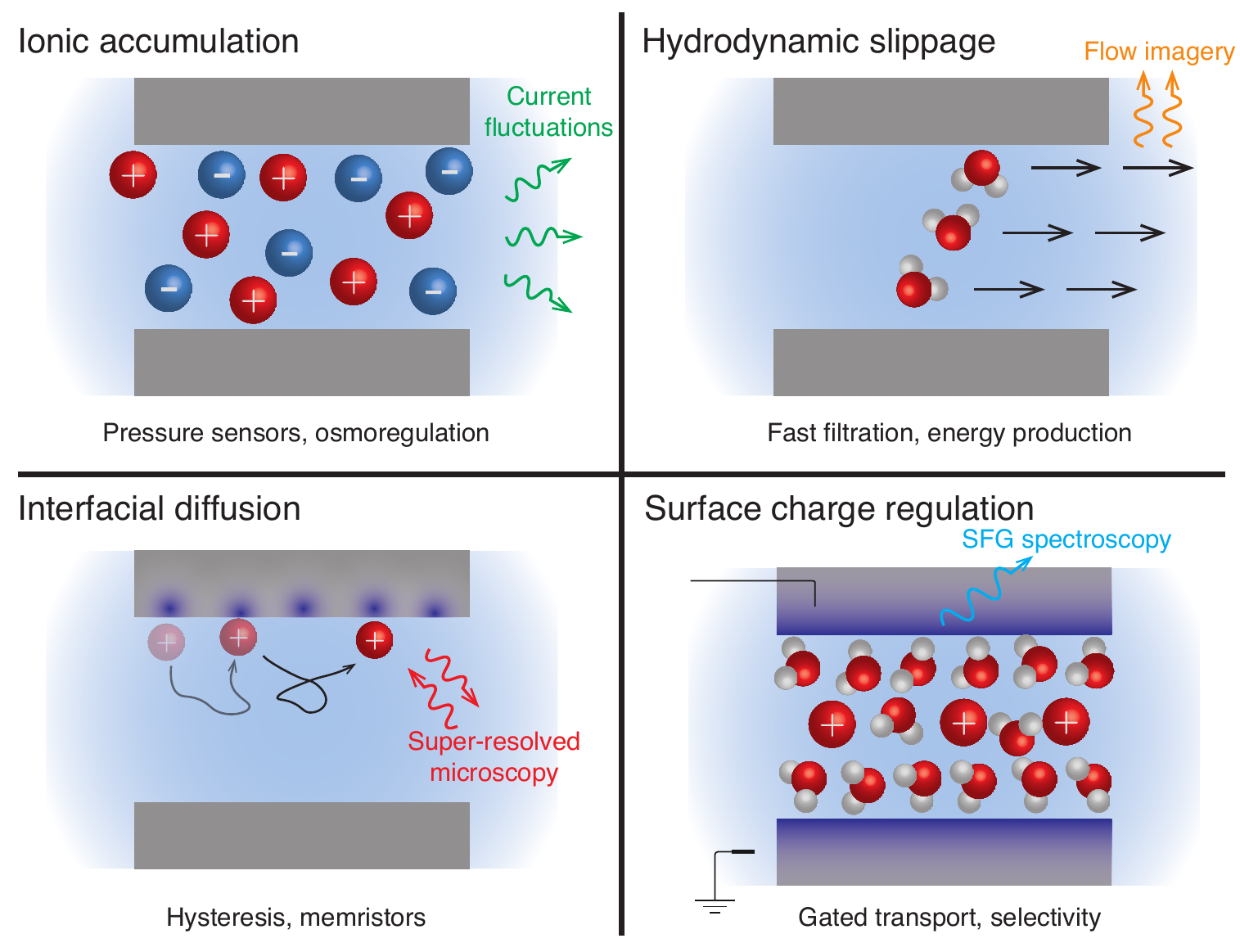}
	\caption{\textbf{Nanofluidics at the crossroads.} Schematic representation of various physical phenomena occuring in nanofluidics, and ways to probe and exploit them.}
\end{figure*}

\LB{Looking back, nanofluidics was born in the footsteps of microfluidics and membrane science. Its original perspectives were relatively modest and merely envisionned phenomena ranging from flows in small channels to separation and sieving of molecular species. But the recent advances in the field make it clear that there is far more to harvest from the properties of fluids at the nanoscale. A whole cabinet of curiosities shows up at the nanoscales. These allow to design advanced ionic machines, hence mimicking some of the most subtle functionalities observed in Nature. This is a challenging quest and, from the observation of the functioning of biological nanofluidic machines, we have a number of key  ingredients to be harnessed in the future:  the breakdown of continuum effects (in both hydro- and  electro- dynamics), far from equilibrium transport phenomena, working at the edge of $k_BT$ and taking benefit from fluctuations, but also quantum engineering of the confining materials. This will allow to embrace the full power of nanofluidic phenomena and recent advances paved the way in this direction, with the design of voltage-gated channels, ion pumps, nanofluidic memristors, etc. as we discussed above.}
	
\LB{So, what would be a roadmap for the development of nanofluidics and the developement of ionic machines?} \LBnew{Such a view is inherently subjective, but we propose height milestones, which could certainly help move the field forward. }\\
\vskip0.1cm
\noindent$\bullet$ \LBnew{\it The excitability-robustness trade-off:}	\LB{As a first general observation,} the properties of biological channels are governed by a complex interplay between chemical, electrical and mechanical information (respectively: neurotransmitters; action potentials; pressure). This complexity is enabled by the fact that all relevant energy scales fall in the range of $0.5-5 \, k_B T$ at nanometric scales, allowing to convert external stimulation into conformational changes in proteins. A notable exception is vision, which detects photons of energy $\sim 1\, \si{eV} \sim 40 \, k_B T$: in humans, photodetection is made possible through cis-trans isomerism of a protein, thus requiring the breaking of a covalent bond (with typical energy $\sim 1\,$eV).

To approach the functionnalities of their biological counterpart, solid-state nanochannels should then aim at exploiting physical processes with typical energy scales around $5\, k_B T$, so that they can be used to mimick the physics of proteins. 
	\LB{In this perspective, optical, chemical, and electronical signals could not only be used as probes, but also as externally controlled knobs for potential ionic machines. This energy scale plays a particular role in the excitability of biological channels -- that is, their ability to react and adapt to even weak stimuli. There appears to be an overall trade-off between robustness and excitability: if the energy barriers separating the different states of a physical device are too large, its behaviour will not be affected by thermal fluctuations; however, it will not be able to detect subtle stimuli, and vice-versa if the barriers are too small. Artificial (e.g. electronic) systems generally attempt to suppress thermal noise by making such barriers as high as possible. The development of ionic machines could be an opportunity to instead exploit fluctuations to reach higher levels of excitability.} \\
\vskip0.1cm
\noindent$\bullet$ \LBnew{\it Stimulating and probing:} The multiplication of possible ways to excite and probe fluidic systems offers leverage for making them interact by expanding the systems' phase space. For example, in a classical fluidic cell where several nanofluidic devices are deposited on the same membrane separating two reservoirs, one can usually only control the electrical potential of the reservoirs. In this situation, all devices are exposed to the same voltage drop, leaving little room for complex interaction. However, one could imagine to use light to specifically activate certain devices, creating an avenue to design advanced ionic machines.  \LBnew{Conversly to such stimuli, a challenge will be the local measurement of properties, {\it e.g.} electrical potentials, ion concentrations, or water dynamics.} All these perspectives are summarized in Fig.~6.\\
\vskip0.1cm
\noindent$\bullet$ \LBnew{\it Nanofluidic machines:} \LB{In terms of circuitry}, we believe that gated channels discussed in Section II \LB{are the building blocks which} offer some of the most promising perspectives \cite{kavokine2019ionic,robin2023long,xiao2019artificial}. They potentially allow to control ion transport at the molecular level, through the use of `switches' on the surface of the pores \rev{-- playing a role similar to transistors in electronics}. One should aim at ascertaining the precise physical mechanisms behind nanofluidic gating, to control them more finely. The development of ionic pumps is a particularly important application of gated channels. Biological pumps are key to maintaining cells out of thermal equilibrium; they also play a major role in the efficiency of water filtration in kidneys \cite{marbach2016active}. Likewise, the subtle dynamics of voltage-gated ion channels underpins many processes in neurotransmission and sensory detection. We believe that artificial ionic pumps and nanofluidic memristors could serve as vital building blocks for artificials proto-cells, i.e. devices that would actively sustain concentration gradients with their environment.\\
\vskip0.1cm
\noindent$\bullet$	\LBnew{\it Many-body effects:} \LBnew{In order to develop advanced transport properties,} one notable aspect that would be worth greater attention is many-body (coulombic) effects. Biological channels are typically very short (1 or 10\,nm in length), so that they will rarely contain more than one ion; changes in their properties can therefore only occur through interactions between single ions and the proteins making up the surface of the pore. The surface of solid-state channels have a much simpler structure than proteins -- however, they easily reach $1\, \si{\micro \meter}$ in length and contain hundreds of ions at any given time. \LBnew{The interaction confinement furthermore increases interactions between charged species in smaller dimensions \cite{kavokine2022interaction}}. As a result, collective behaviours, like ionic pairing \LBnew{or ionic Coulomb blockade}, are likely to play a significant role in their transport properties -- which could in turn be used to reach complex functionalities.\\
\vskip0.1cm
\noindent$\bullet$ \LBnew{\it Iontronics and neuromorphic mimicry:} \LB{The recent \PR{development of ion-based, nanofluidic memristors paves the way for `iontronics' -- transporting, processing and storing information through the use of solvated ions in water \cite{robin2023long,xiong2023neuromorphic,noy2023nanofluidic}}. These experiments already demonstrated the possibility of implementing some form of Hebbian learning \cite{robin2023long}, a \PR{basic form of learning algorithm}. Looking at the brain performance, the advantage \PR{of using ions rather than electrons seems to be} manifold: lower energy consumption, hardware-level plasticity, and multiple information carriers \cite{noy2023nanofluidic}, as well as the possibility of bio-artificial ionic interconnections. We therefore have building blocks at our disposal to create a future neuromorphic computer based on ions as charged carriers. There is still a long way to reach this goal, and this requires to invent and build a nanofluidic circuitry.} \PR{However, exploring the potential for a form of `ionic supremacy' would be a worthwhile goal. 
}\\
\vskip0.1cm
\noindent$\bullet$ \LBnew{\it Iontronic circuitry:}  \LB{Building fluidic networks is a key challenge to be adressed}. As of now, integrating nanofluidic systems into fluidic networks \LB{has not been achieved}. Existing devices are generally characterized in transmembrane setups, where the device is deposited on a membrane separating two controlled reservoirs. Electrodes immersed in each of the reservoirs allow to control the voltage drop imposed on the device. 
This geometry leaves only little room for studying interacting devices beyond arranging them in parallel on the same membrane. In particular, it limits measures to global quantities like the total device conductance. The development of more local probes, such as in situ electrodes, would grant access and control to the microscopic properties of nanofluidic systems. In addition, the development of van der Waals assembly offers new way to design a nanofluidic circuitry \cite{geim2013van}.
 \PR{Another promising approach is that of carbon nanoporins\cite{tunuguntla2017enhanced} -- very short, subnanometer carbon nanotubes that are known to spontaneously insert in lipidic membranes. These channels display rich physical properties like voltage gating \cite{yao2021electrostatic,li2022breakdown}, and could be used to create ionic machines with cell-like design.}\\
%
%
%
\vskip0.1cm
\noindent$\bullet$ \LBnew{{\it Quantum engineering of flows and fluxes:} The coupling of flows to the quantum properties of the confining materials is an asset which has been barely explored. 
 Going down this road requires a fundamental understanding of the properties of water and its coupling to the environment, down to the smallest, angstr\"om scale confinement. 
 This is the frontier where the continuum of fluid mechanics meets the atomic or even quantum nature of matter \cite{kavokine2022fluctuation,lizee2023strong}. A prerequisite is to 'see' the molecular dynamics of water in such ultimate corridors, and powerful experimental techniques like SFG and dielectric spectroscopy should be decisive to advance on these questions \cite{chiang2022dielectric}. In parallel, it becomes necessary to develop dedicated theoretical framework at the interface between soft and hard condensed matter, in order to account for the emerging properties of liquids at these scales \cite{kapil2022first,kavokine2022fluctuation,coquinot2023quantum}. Harvesting properly the electronic properties of the confining materials would allow to develop a 'quantum engineering' of nanofluidic flows. Incidentally, this would open manifold possibilities to exploit novel iontronic-electronic couplings.}
\\
\vskip0.1cm
\noindent$\bullet$ \LBnew{\it Nanoscale robotics:} \LB{Last but not least, there are direct connections to the field of nano-robotics. The recent demonstration of a nanopore-powered DNA turbine \cite{shi2022nanopore}, built on the basis of  DNA origami technology and powered by electro-chemical gradients, opens further the phase space of possibilties. Such device allows to transduce local free energy into mechanical motion. This therefore connects nanoscale fluid transport to mechanical work.}

Nanofluidics now sits at the crossroads: important advances are within reach, but paths leading to them involve concepts that extend beyond traditionnal \LB{methodologies}. The field as the whole approaches a watershed as new techniques can provide unprecedented microscopic information on nanoscale physics, and can in turn transform nanofluidic devices into polyvalent ionic machines. 

\PR{As Feynman famously stated, there is plenty of room at the bottom. This phrase definitely found its echo in solid-state physics, and it is now the turn of soft matter scientists to risk venturing there -- and catch all the weird fishes roaming molecular-scale waters.}
	
	
	\begin{acknowledgments}
	The authors thank Nikita Kavokine for interesting discussions.
	L.B. acknowledges funding from the EU H2020 Framework Programme/ERC Advanced Grant agreement number 785911-Shadoks. 
	\end{acknowledgments}

	\section*{Data Availability Statement}
	Data sharing is not applicable to this article as no new data were created or analyzed in this study.
	
	\bibliography{perspbib}

	\end{document}